\documentclass[twocolumn,prl,aps,superscriptaddress, longbibliography]{revtex4-1}
\usepackage[T1]{fontenc}
\usepackage[latin9]{inputenc}

\usepackage{CJK}

\usepackage{amsmath}
\usepackage{amssymb}
\usepackage[usenames,dvipsnames]{xcolor}
\usepackage{bbm}
\usepackage{braket}
\usepackage{mathrsfs}  
\usepackage{mathtools}
\usepackage{physics}

\allowdisplaybreaks

\usepackage{comment}

\usepackage{graphicx}

\usepackage{textcmds}

\usepackage[normalem]{ulem}

\usepackage[colorlinks=true]{hyperref}  
\hypersetup{
    bookmarks=true,         
    unicode=false,          
    pdftoolbar=true,        
    pdfmenubar=true,        
    pdffitwindow=false,     
    pdfstartview={FitH},    
    pdftitle={Euler classical spin liquid},    
    pdfauthor={Ruegg, Morris, Yan, Slager},     
    pdfsubject={},   
    pdfcreator={},   
    pdfproducer={}, 
    pdfkeywords={} {} {}, 
    pdfnewwindow=true,      
    colorlinks=true,       
    linkcolor=blue, 
    citecolor=blue,        
    filecolor=magenta,      
    urlcolor=blue,           
	breaklinks=true
} 

\definecolor{orange}{rgb}{1,0.5,0}

\newcommand{\sect}[1]{\vspace{0.3em}{\it #1.}---}

\usepackage{bbold}

\newcommand{\be}{\begin{equation}}
\newcommand{\bea}{\begin{eqnarray}}
\newcommand{\eea}{\end{eqnarray}}

\newcommand{\ii}{\mathrm{i}}
\newcommand{\ee}{\mathrm{e}}

\renewcommand{\aa}{\mathrm{a}}
\newcommand{\Eu}{\mathrm{Eu}}

\begin{document}

\begin{CJK*}{UTF8}{gbsn} 

\author{Luca R\"uegg}
\email{lr537@cam.ac.uk}
\address{TCM Group, Cavendish Laboratory, University of Cambridge, J. J. Thomson Avenue, Cambridge CB3 0US, United Kingdom}
\author{Arthur Morris}
\address{TCM Group, Cavendish Laboratory, University of Cambridge, J. J. Thomson Avenue, Cambridge CB3 0US, United Kingdom}
\author{Han Yan (闫寒)} 
\affiliation{Institute for Solid State Physics, The University of Tokyo,  Kashiwa, Chiba 277-8581, Japan} 
\author{Robert-Jan Slager}
\email{rjs269@cam.ac.uk}
\affiliation{Department of Physics and Astronomy, University of Manchester, Oxford Road, Manchester M13 9PL, United kingdom}
\address{TCM Group, Cavendish Laboratory, University of Cambridge, J. J. Thomson Avenue, Cambridge CB3 0US, United Kingdom}

\title{Euler topology in classical spin liquids}

\date{\today}
\begin{abstract} 
Classical spin liquids have recently been analyzed in view of the single-gap homotopy classification of their dispersive eigenvectors.
We show that the recent progress in defining multi-gap topologies, notably exemplified by the Euler class, can be naturally included in these homotopy-based classification schemes and present phases that change topology by band node braiding.
This process alters the topology of the pinch points in the spin structure factor and consequently their stability. Furthermore, we discuss how these notions also pertain to models discussed previously in the literature and have a broader range of application beyond our specific results. Our work thus opens up an uncharted avenue in the understanding of spin liquids.
\end{abstract}

\maketitle

\end{CJK*}

\sect{Introduction}
Classical spin liquids (CSLs) are spin models that possess an extensive ground state degeneracy which prevents ordering at the lowest temperatures, representing a paradigmatic example of a frustrated system.
In addition, they serve as parent systems for quantum spin liquids (QSLs), for example, quantum spin ice \cite{Bramwell2001,Castelnovo2008,Gingras2014}. Several classifications of QSLs have been known for some time, such as those based on projective symmetry groups~\cite{Wen2002} or topological quantum field theories~\cite{Essin2013,Barkeshli2019}.

Recently, classification schemes for CSLs with continuous spins have been proposed~\cite{Yan2023typology,Yan2023detailed,Yan2024PhysRevB,Davier2023} in which one can exploit the spectra of the diagonalized interaction Hamiltonians,  providing a route to translate band theory insights from tight-binding Hamiltonians to CSLs.
More specifically, CSLs have at least one flat band at zero energy encoding the extensive ground state degeneracy, and dispersive bands above for excited states. 
Subsequently, a distinction can be made between fragile topological CSLs (FT-CSLs) which have a gap between the flat and dispersive bands, and algebraic CSLs (A-CSLs) which do not~\cite{Yan2023typology}.
Assuming a partitioning of the bands into the ground state manifold and gapped dispersive bands finally allows for a classification of FT-CSLs~\cite{Rehn2017,fancelli2024fragilespinliquiddimensions} analogous to that of topological band theories~\cite{Rmp1, Rmp2, Armitage_2018}.
Similarly to fragile topological insulators, the topology of FT-CSLs can be trivialized by adding trivial bands~\cite{ft1, bouhon2018wilson}; this is, however, less of an issue for spin liquids where the number of spins per unit cell is usually well controlled.
A-CSLs exist as the  phase boundaries between different FT-CSLs and exhibit pinch point singularities in the spin structure factor at the gap-closing points.
The Taylor expansion of the eigenvectors around the band touching points in the A-CSLs determines the form of the generalized Gauss's law dominating their low-energy physics \cite{Benton2016,PhysRevLett.124.127203,PhysRevB.98.165140,PhysRevLett.130.196601,Benton2021PhysRevLett,Placke2024PhysRevB}.

Independently, recent developments in band topology have led to the discovery of new multi-gap phases in which the (fragile) homotopy structure of the total spectrum is important, as opposed to the single-band topology alone; these multi-band phases do not fit into existing classifications~\cite{clas1,SchnyderClass,rjs_translational, Slager_NatPhys_2013, Kitaev20062, Kruthoff_2017,Clas4, Clas5}. In such systems, {\it groups of bands}, or band subspaces, acquire a topological character~\cite{Bouhon2020_geo,bouhon2018wilson, bouhon2022multigap, davoyan2023mathcalpmathcaltsymmetric,bouhon2023quantum,bouhon2022multigap, Lim2023, PhysRevB.110.075135}. A prime example is the Euler class~\cite{Bouhon_2019,PhysRevX.9.021013, PhysRevLett.118.056401} - the real analogue of the Chern number - in which two-band subspaces give rise to an invariant that is manifested by the presence of nodes that cannot be removed. The nodes between bands host frame charges which can only be altered (to induce a change of Euler class) when braided around nodes residing between the next pair of bands~\cite{, Bouhon_2019,PhysRevX.9.021013,   bouhon2018wilson, Bouhon2020_geo, ahnprl}. This is possible as the frame charges of nodes between two sets of adjacent two-band subspaces (the pairs of bands) are non-Abelian~\cite{doi:10.1126/science.aau8740}. The observation and manipulation of these novel topological phases presents an increasingly active research field and has been notably related to out-of-equilibrium quenches and Floquet systems~\cite{Unal_2020,slager2024floquet,zhao2022observation,breach2024interferometry}, phonon spectra~\cite{Peng2021} and magnetic systems~\cite{mag2024euler,magnetic}, as well as metamaterials~\cite{Jiang_2021, natcom4band,JIANG2024, Guo1Dexp}, and recently even interacting phases~\cite{Wahl2025Peps}.

Here, we apply these novel insights in multi-gap topological phases to CSLs. Notably, our findings challenge the common view that the low-energy effective physics is determined only by the bottom part of the spectrum that includes the bottom flat bands and the dispersive band around the neighborhood of the gapless points; while the higher energy bands, whose states are not excited, are considered irrelevant. Indeed, we show that the full homotopy structure in fact can be of relevance as quantified by the appearance of multi-gap invariants. To concretize the discussion we present our findings in terms of a tractable two-dimensional three-band Heisenberg model with one bottom flat band that has nodes with finite Euler class. Upon braiding nodes between the two upper dispersive bands around the Euler nodes, the sign of the Euler class switches. As a consequence, the mutual stability of the Euler nodes is affected, which is reflected in the pinch-point topology. We also find that the emergent Gauss's law remains invariant and the pinch points themselves do not change their $n$-foldness. We then use these insights to formalise our general findings, discuss the natural role of multi-gap topology in CSLs, and show that several models in earlier literature can be understood in terms of the Euler topology of the flat bands.

\sect{Classification of CSLs}
Many classical spin models can be accurately captured using the soft-spin approximation, where the local hard-spin constraint $|\vb{S}(\vb r)| = 1$ is relaxed to the averaged constraint over all sites $\langle \vb{S}^2(\vb r) \rangle = 1$. This approximation has the effect of decoupling the individual spin components, meaning that they can each be treated as real scalars.
A classical spin model with bilinear interactions can then be written in momentum space as
\begin{equation}
    H = \frac{1}{2} \sum_{\vb k} \vb {\tilde{S}}(\vb k)^{\intercal} J(\vb k) \vb {\tilde{S}}(\vb k),
\end{equation}
where $\vb {\tilde{S}}  (\vb k)$ is the Fourier-transformed vector of the $N$ scalar spins in a unit cell.
The interaction matrix $J(\vb k)$ is positive-semidefinite and of rank $M$.
The spectrum is described by $M$ dispersive higher bands, whose eigenvectors are $\{ \vb T_I(\vb k) \}_{I=1,\dots,M}$, and $N-M$ bottom flat bands, with eigenvectors in the orthogonal space spanned by $\{ \vb B_J(\vb k) \}_{J=1,\dots,N-M}$.

The classification of FT-CSLs in Ref.~\cite{Yan2023typology} is based on topologically distinct mappings from the $d$-dimensional Brillouin zone $\mathbb{T}^d$ to the classifying space $G$, defined as the space spanned by the eigenvectors of the $M$ highest bands (or equivalently the $N-M$ lowest bands).
The equivalence class of these maps is the the homotopy class $[\mathbb{T}^d, G]$.
As a concrete example, the Kagome-star model~\cite{Rehn2017,Yan2023typology,Davier2023}, whose interaction matrix is real and has rank 1, has the classifying space of the real Grassmannian $\text{Gr}_{2,3}^{\mathbb{R}} = O(3) / (O(2) \times O(1)) = \mathbb{RP}^2$.
Provided the Berry phases around the non-trivial loops of the Brillouin zone vanish, the homotopy class is equivalent to $\pi_2 (\mathbb{RP}^2) = \mathbb{Z}$.
Hence, the FT-CSL phases of the Kagome-star model are labeled by an integer number, which turns out to be equivalent to the Skyrmion number of the dispersive band eigenvector, or equivalently, the Euler class of the flat bands.

\sect{Multi-gap classification}
The key insight of multi-gap classifications~~\cite{Bouhon2020_geo,bouhon2018wilson, bouhon2022multigap, Bouhon_2019,davoyan2023mathcalpmathcaltsymmetric,bouhon2023quantum,bouhon2022multigap, Lim2023} is that upon considering finer partitions of the spectrum, new homotopy classifications arise. For example, in the case of a real two-dimensional three-band model where all bands are gapped from one another, the classifying space is the Flag manifold $\text{Fl}_{1,1,1}^{\mathbb{R}} = O(3) / (O(1) \times O(1) \times O(1))$, coinciding with the configuration space of a biaxial nematic $SO(3) / D_2$~\cite{Kamienrmp, Liu2016prx, volovik2018investigation, Beekman20171}. The charges of nodes are set by the first homotopy group $\pi_1 (SO(3) / D_2) = \mathbb{Q}$, corresponding to vortices in the frame of eigenvectors $\{ \vb u_i(\vb k) \}_{i=1}^3$ labeled by the conjugacy classes that are the Quaternions $\{ \pm 1, \pm \ii, \pm \mathrm{j}, \pm \mathrm{k} \}$.
Braiding nodes in different gaps around each other changes the sign of their charge, in other words they anti-commute. Thus, braiding processes can change the relative stability of nodes between bands $\ket{u_{1,2}(\vb k)}$. 

The mutual stability of nodes due to braiding may be analyzed through the (patch) Euler class. 
This `local' topological invariant, defined for a domain $\mathcal{D}$ in the Brillouin zone (BZ) as
\begin{equation}
    \chi(\mathcal{D}) = \frac{1}{2\pi} \left[ \int_{\mathcal{D}} \Eu - \int_{\partial\mathcal{D}} \aa\right],
\end{equation}
is well-defined whenever the two bands $\ket{u_{1, 2}}$ do not have nodes with any other bands in $\mathcal{D}$. Here  $\aa(\vb k) = \braket{u_1}{\dd u_2}$ is the Euler connection, while $\text{Eu}(\vb k) = \dd \, \aa(\vb{k})$ is the Euler form.
The patch Euler class is an integer whenever $\mathcal{D}$ contains an even number of nodes, and when nonzero it indicates 
that the nodes cannot annihilate. In the example described above, the Euler class $\chi=\chi(\text{BZ}=\mathbb{T}^2)$ of a two-band subspace in a three-band real Hamiltonian, which is equivalent to the invariant $\pi_2 (\mathbb{R}P^2) = \mathbb{Z}$, changes by an integer after braiding has been carried out.

\sect{Euler CSL model}
We now formulate a three-site square-lattice Heisenberg CSL model that possesses finite Euler class nodes and showcases non-Abelian braiding.
The starting point is the two orthogonal vectors $\vb v_m^3(\vb k) = 2 \, (m + \cos k_x, \cos \frac{k_x - k_y}{2}, \cos \frac{k_x + k_y}{2})^{\intercal}$ and $\vb v^2(\vb k) = \frac{1}{2} \, (0, - \cos \frac{k_x + k_y}{2}, \cos \frac{k_x - k_y}{2})^{\intercal}$ which we use to define the interaction matrix as~\cite{supplementary}
\begin{equation}\label{eq:Hamiltonian_m}
    J_m(\vb k) = \vb v_m^3(\vb k) \, \vb v_m^3(\vb k)^{\intercal} + \vb v^2(\vb k) \, \vb v^2(\vb k)^{\intercal}.
\end{equation}
Here, $m$ is a tuning parameter. 
This Hamiltonian is real and thus manifestly possesses $\mathcal{C}_2 \mathcal{T}$ symmetry.
The spectrum features one flat band at zero energy because for all $m$ the Hamiltonian has rank 2 or less.
The two dispersive bands lie above the flat one since the Hamiltonian is explicitly positive semidefinite.
The second band touches the flat one at the high-symmetry points $\pm X$ and $\pm Y$ for all $m$, where $X=(\pi,0)$ and $Y = (0, \pi)$.
The third band is gapped from the second one except for $m = +1$ [$m=-1$] when it touches at $\pm X$ [resp. $\pm Y$].

\begin{figure}
\flushleft
\includegraphics[width=0.5\textwidth]{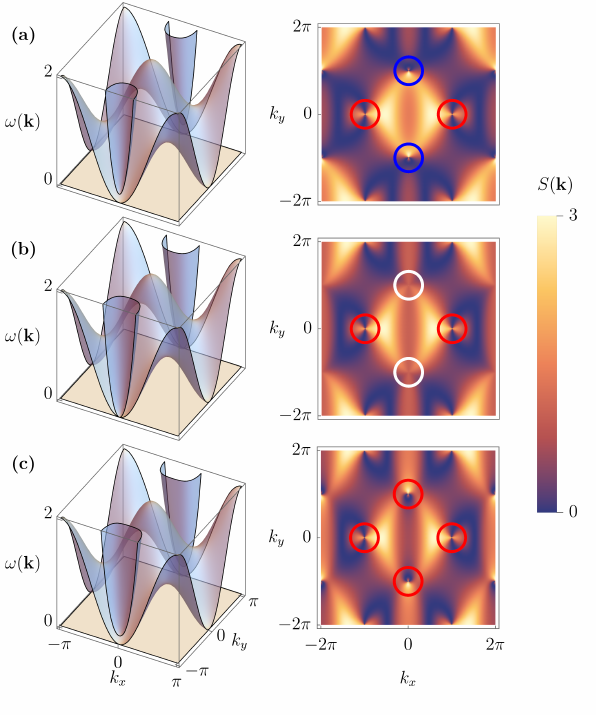}
\caption{Spectrum and spin structure factor when \textbf{(a)} $m=-1.2$ \textbf{(b)} $m=-1$ and \textbf{(c)} $m=-0.8$. Coloured rings in the structure factor indicate the patch Euler class of the node: red [blue] indicates $\chi(\mathcal{D})=+1$ [$\chi(\mathcal{D})=-1$], while nodes where the top band touches the bottom band are shown in white}
\label{fig: spectrum and Sq Euler nodes}
\end{figure}

When the third band is gapped the Euler class for the lower two bands is well-defined, as is the patch Euler class of the all the nodes between the first and second bands.
All the nodes possess a finite patch Euler class of $\pm1$ indicating that they are all stable Euler nodes, meaning they cannot be removed by small symmetry-preserving perturbations.
The sign of the Euler class only has a physical significance when considering the mutual stability of Euler nodes: nodes of  opposite [equal] sign can [cannot] annihilate.
The phase diagram as a function of $m$ is as follows: for $m < -1$ the Euler class is $\chi = 0$; at $m = -1$ the band touching at $\pm Y$ changes the sign of the corresponding Euler nodes (see Fig.~\ref{fig: spectrum and Sq Euler nodes}), which implies that $\chi = 2$ for $-1 < m < 1$; and the band touching at $\pm X$ which occurs at $m=1$ likewise changes the sign of the corresponding Euler nodes, leading to $\chi = 0$ for $m>1$. Gap closings (such as the one at $m=-1$) are necessary for topological invariants such as $\chi$ to change. These transitions may be observed in the spin structure factor at zero temperature, 
\begin{equation}
    S(\vb{k})= \left| \sum_{i=1}^3 (\hat{v}_m^1)^i (\vb k) \right|,
\end{equation}
where $\hat{\vb{v}}_m^1 (\vb k) = \vb v^2 (\vb k) \times \vb v_m^3 (\vb k)/|\vb v^2 (\vb k) \times \vb v_m^3 (\vb k)|$ is the flat band eigenvector.  
In particular, the pinch points, singularities of the structure factor which act as signatures of higher bands touching the flat band and determine the associated emergent Gauss's laws, are modified trivially at the gap-closing points (see Fig.~\ref{fig: spectrum and Sq Euler nodes}). We discuss this further in Appendix \ref{sec:Gauss}.

\begin{figure}
\centering
\includegraphics[width=0.48\textwidth]{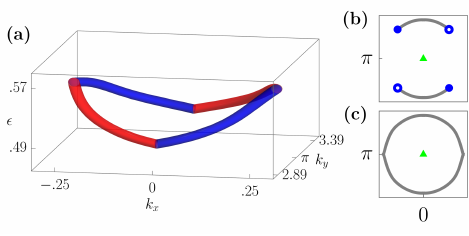}
\caption{\textbf{(a)} Illustration of the braiding process realised at the $Y$ point due to the coupling in Eq.~(\ref{eq:BraidingCoupling}); see main text for details. \textbf{(b)} Dirac strings attached to the nodes keep track of gauge discontinuities. \textbf{(c)} After the nodes have been annihilated, the Dirac strings can only be removed by moving them past the node at the $Y$ point, thereby flipping its charge.}
\label{fig:Braiding3D}
\end{figure}

The crucial insight from the non-Abelian braiding of band nodes known from topological band theory is that the sign of the Euler nodes in one gap can be changed by braiding Weyl nodes in adjacent gaps around it.
In this way, the pinch points should persist while we change the patch Euler class, and consequentially the relative stability of the pinch points.
Consider the vectors $\vb w_{\epsilon}^3(\vb k) = (1-\epsilon) \, \vb v_{m = -1.2}^3(\vb k) + \epsilon \, \vb v^2(\vb k)$ and $\vb w_{\epsilon}^2(\vb k) = (1-\epsilon) \, \vb v^2(\vb k) + \epsilon \, \vb v_{m = -0.8}^3(\vb k)$, and define
\begin{equation}\label{eq:BraidingCoupling}
    \tilde{J}_{\epsilon}(\vb k) = \vb w_{\epsilon}^3(\vb k) \, \vb w_{\epsilon}^3(\vb k)^{\intercal} + \vb w_{\epsilon}^2(\vb k) \, \vb w_{\epsilon}^2(\vb k)^{\intercal}.
\end{equation}
This interaction matrix interpolates between $J_{m = -1.2}(\vb k) = \tilde J_{\epsilon = 0}(\vb k)$ and $J_{m = -0.8}(\vb k) = \tilde J_{\epsilon = 1}(\vb k)$.
It has rank 2 everywhere except the high-symmetry points where it has rank 1 for all $\epsilon$, so the pinch points remain invariant.
Approaching $\epsilon = 0.49$, two pairs of Weyl nodes between bands 2 and 3 form in the vicinity of the $\pm Y$ point.
Subsequently, the nodes of each pair move apart from one another and towards a node of the other pair, with which they eventually annihilate at $\epsilon \approx 0.57$ (see Fig.~\ref{fig:Braiding3D}).
Upon creation, the two Weyl nodes in a given pair have opposite frame charges and are connected by a Dirac string - an unphysical line that keeps track of the gauge obstruction. We are free to place this string anywhere, provided it connects two Weyl nodes.
As the nodes are annihilated in different pairs to those in which they were created, it is necessary for the Dirac string to pass through the Euler node at the $\pm Y$ point during the braiding process.
The patch Euler class then changes sign, causing the total Euler class to jump from 0 to 2 (this latter quantity is well-defined only after the nodes have annihilated).

\begin{figure*}
\centering
\includegraphics[width=\textwidth]{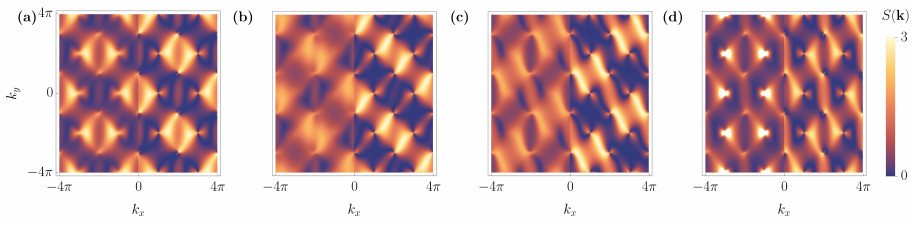}
\caption{Spin structure factors corresponding to the model defined by Eq.~(\ref{eq:BraidingCoupling}) for \textbf{(a)} $\epsilon = 0.2$ \textbf{(b)} $\epsilon = 0.4$ \textbf{(c)} $\epsilon = 0.6$ and \textbf{(d)} $\epsilon = 0.8$. The left half is calculated with Monte Carlo simulations at $T = 0.1$, whereas the right half is based on the zero-temperature case within Luttinger-Tisza.}
\label{fig:Sqhalfhalf}
\end{figure*}

\sect{Monte Carlo calculations}
To ensure that our model exhibits a spin liquid phase beyond the analytical calculation within the soft-spin approximation at zero temperature, we employ finite-temperature Monte Carlo calculations in a parallel tempering scheme.
The Markov chain is initialized in one of two types of ground states that we understand~\cite{supplementary}, or a random state. Before the measurement phase, the system is thermalized in a simulated annealing and subsequent constant-temperature procedure.
To update the Markov chain we use a combination of Metropolis, overrelaxation, and gauge steps.
The size of our system is $L \times L$ lattice sites with $L = 64$.
We perform $10^4$ annealing, relaxation, and sampling sweeps to calculate the spin structure factor
\begin{equation} \label{eq:SpinStructureFact}
    S(\vb{k})= \sqrt{\Big\langle \Big|\sum_j \vb S(\vb r_j) \, e^{i \vb k \cdot \vb r_j}\Big|^2 \Big\rangle},
\end{equation}
where $\langle\cdot\rangle$ denotes the thermal expectation value.
The temperature is always expressed in units of $k_{\text{B}}$.
Around $T = 0.1$, the MC spin structure factor of $\tilde J_{\epsilon} (\vb k)$ shows distinct pinch points (see Fig.~\ref{fig:Sqhalfhalf}), a telltale spin liquid feature, for all $0 \leq \epsilon \leq 1$.
It also agrees in good qualitative terms with the analytically calculated zero-temperature structure factor.

\begin{figure}
\flushleft
\includegraphics[width=0.45\textwidth]{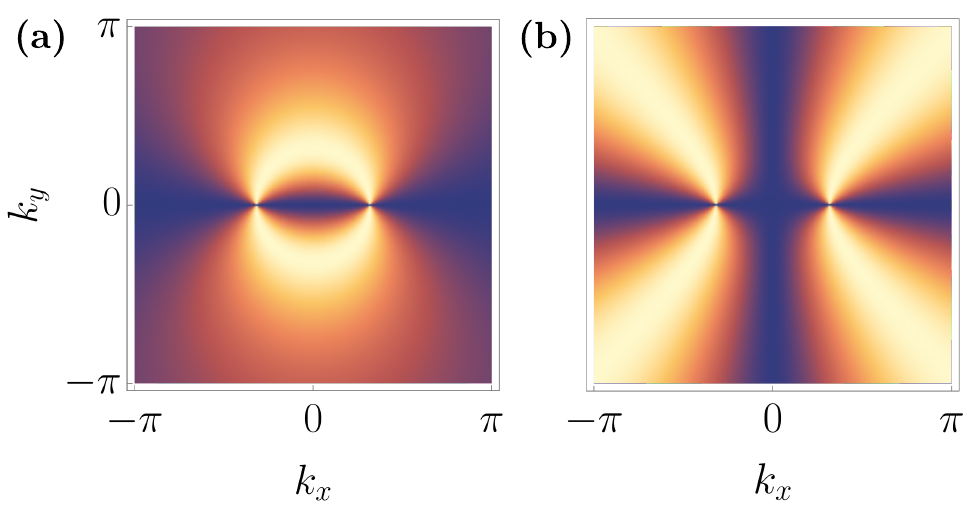}
\caption{Characteristic  spin structure factors of a pair of pinch point nodes when their total patch Euler class is \textbf{(a)} zero ($i=1$ in Eq.~(\ref{eq:2BandModel})) and \textbf{(b)} non-zero ($i=2$ in Eq.~(\ref{eq:2BandModel})).}
\label{fig:2BandModelStructureFactors}
\end{figure}

\sect{Pinch-point structure}
While the Euler class can only exist in models with three or more bands~\cite{Bouhon_2019, Bouhon2020_geo}, the characteristic structure of pinch points of Euler nodes can be captured by a simple two-band model. Setting $z=k_x+\ii k_y$, $Z_1 = (\bar{z}+m)^2(z-m)^2$, and $Z_2 = (z+m)^2(z-m)^2$ we define a pair of real $2\times 2$ Hamiltonians for $i=1, 2$, as
\begin{equation}\label{eq:2BandModel}
    h_i(\vb{k}) =Z_i\sigma_- +\bar{Z}_i\sigma_+
\end{equation}
where $\sigma_{\pm}=(\sigma_x\pm\ii\sigma_z)/2$. For $m\neq 0$ both $h_1(\vb{k})$ and $h_2(\vb{k})$ have nodes at $\vb{k}=(\pm m, 0)$. However, the topological character of these nodes differs in each case: for $i=1$ the nodes carry opposite winding numbers $\pm 1$, whereas for $i=2$ the charges are equal. This difference is reflected in the spin structure factors (see Fig. \ref{fig:2BandModelStructureFactors}), which display features similar to the three-band examples. 

\sect{Kagome-star model}
The understanding of multi-gap topology also allows for a reinterpretation of known CSLs.
We will exemplify this with the Kagome-star model, a CSL on the Kagome lattice with one constrainer~\cite{Rehn2017,Yan2023typology,Davier2023}, meaning that there are two flat bands at zero energy and one dispersive band above.
In its classification diagram, which is controlled by two interaction coefficients, there are several lines of A-CSL phases between patches of FT-CSLs.
The different FT-CSL phases are labeled by the winding number of the dispersive eigenvector around the Brillouin zone torus, which is also called the Skyrmion number $Q_{\text{Sk}}$.

In the case where the two bottom bands are gapped from the top band, the Euler class coincides with the Skyrmion number.
This means that the FT-CSL phases in the Kagome-star model have a finite Euler class.
Calculating the flat eigenvectors shows that they possess singularities connected by Dirac strings.
Because the two bottom bands are flat, however, these singularities and Dirac strings are completely gauge-dependent and do not show up in physical quantities like the spin structure factor.
Only if one were to lift the degeneracy of the two bottom bands, like is the case in our model above, would these singularities become meaningful.

It is interesting though to observe what happens with these singularities and Dirac strings as one goes across transitions between different FT-CSLs.
For example, following the transition from $Q_{\text{Sk}} = 2$ to $Q_{\text{Sk}} = -2$, we see that the singularities merge in the A-CSL at the gap-closing point and then reemerge in the FT-CSL.
The sign change across the Dirac string is inverted, hence the change in sign of the Skyrmion number.
In contrast, following the transition from $Q_{\text{Sk}} = 2$ to $Q_{\text{Sk}} = 6$, two new singularities are created along an existing Dirac string.
This also the reason why the Skyrmion number only changes by multiples of 4.

\sect{Discussion and conclusion}
Recent band-topology-based classifications of CSLs introduced a tool to study these exotic systems.
This allowed us to apply recent results from multi-gap topological band theories in a new context.
We formulated a three-orbital square-lattice Heisenberg spin model possessing $\mathcal{C}_2 \mathcal{T}$-symmetry using the constrainer formalism.
Analyzing its interaction matrix in a Luttinger-Tisza approximation, we recovered the topological features known from $\mathcal{C}_2 \mathcal{T}$-symmetric two-dimensional band models, such as the non-Abelian charges of band nodes.
In the context of spin liquids this is manifested as stable pinch points in the spin structure factor.
We further devised an interpolation model in which the mutual stability of pinch points changes without altering the Gauss's laws of the pinch points.
This stability is induced by braiding nodes between the two dispersive bands around the pinch point.
Crucially, this means that higher dispersive bands can change the topology of the ground state manifold, a previously unrecognized possibility for classical spin liquids. Although our model is only a proof of concept, we expect this phenomenon to arise in many more natural CSL models; indeed, we have pointed out that certain known models in the literature actually have an Euler class.

Finally, to confirm the validity of our results beyond the Luttinger-Tisza approximation, we used state-of-the-art Monte Carlo simulations to calculate the spin structure factor, and found good agreement with the analytical calculations.  While the range of stability of the liquid phase is finite for $\mathsf{O}(3)$ spins, we found that it is enlarged in $\mathsf{O}(N)$ spin models for $N>3$. Moreover, regions of stability are more ubiquitous in this case.
We will report on a full quantitative study of the phase diagram, including potential ordering transitions at lower temperatures, in the near future.

Future prospects include finding realistic implementations of our models as well as using the multi-gap point of view to discover more exotic kinds of topology.
We reiterate that recent progress in band theory can help inform studies of spin liquids. 
For example, the extension of these findings to quantum spin liquids may prove interesting, especially given the natural connection with lattice gauge theories; the nodes of the Euler class are counterparts of $\pi$-vortices of a nematic that can be formulated as a gauge theory~\cite{Liu2016prx, Beekman20171}.
We believe that we have only scratched the surface of the range of new insights into spin liquids enabled by our understanding of multi-gap topological phases, setting a rich agenda for future pursuits.

\begin{acknowledgements}
\sect{Acknowledgements} 
L.~R. cordially thanks Alaric Sanders and Michael Rutter for answering many questions regarding the Monte Carlo implementations.
L.~R. and R.-J.~S acknowledge funding provided by the Winton programme and the Schiff foundation.
A.S.M.~acknowledges funding from EPSRC PhD studentship (Project reference 2606546), and from a Summer Fellowship from the Japan Society for the Promotion of Science. R.-J.S. acknowledges funding from an EPSRC ERC underwrite grant  EP/X025829/1, and a Royal Society exchange grant IES/R1/221060 as well as Trinity College, Cambridge.
H.Y. is supported by the 2024 Toyota Riken Scholar Program from the Toyota Physical 
and Chemical Research Institute, and the  Grant-in-Aid for Research Activity Start-up from the Japan Society for the Promotion of Science (Grant No. 24K22856).
\end{acknowledgements}

\bibliography{bibliography}

\clearpage
\onecolumngrid
\appendix

\renewcommand\thefigure{\thesection.\arabic{figure}}



\section{Real-space model}
\setcounter{figure}{0}

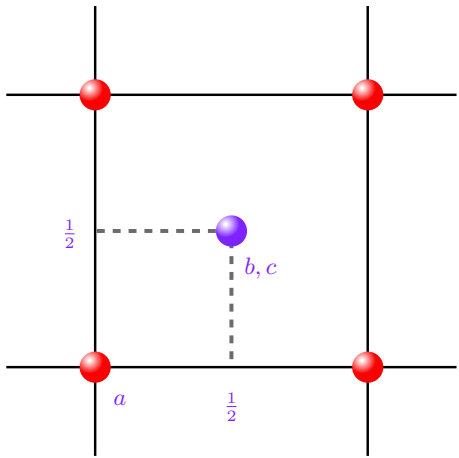
\begin{figure}
\centering
\begingroup%
  \makeatletter%
  \providecommand\color[2][]{%
    \errmessage{(Inkscape) Color is used for the text in Inkscape, but the package 'color.sty' is not loaded}%
    \renewcommand\color[2][]{}%
  }%
  \providecommand\transparent[1]{%
    \errmessage{(Inkscape) Transparency is used (non-zero) for the text in Inkscape, but the package 'transparent.sty' is not loaded}%
    \renewcommand\transparent[1]{}%
  }%
  \providecommand\rotatebox[2]{#2}%
  \newcommand*\fsize{\dimexpr\f@size pt\relax}%
  \newcommand*\lineheight[1]{\fontsize{\fsize}{#1\fsize}\selectfont}%
  \ifx\svgwidth\undefined%
    \setlength{\unitlength}{169.65424185bp}%
    \ifx\svgscale\undefined%
      \relax%
    \else%
      \setlength{\unitlength}{\unitlength * \real{\svgscale}}%
    \fi%
  \else%
    \setlength{\unitlength}{\svgwidth}%
  \fi%
  \global\let\svgwidth\undefined%
  \global\let\svgscale\undefined%
  \makeatother%
  \begin{picture}(1,1)%
    \lineheight{1}%
    \setlength\tabcolsep{0pt}%
    \put(0,0){\includegraphics[width=\unitlength,page=1]{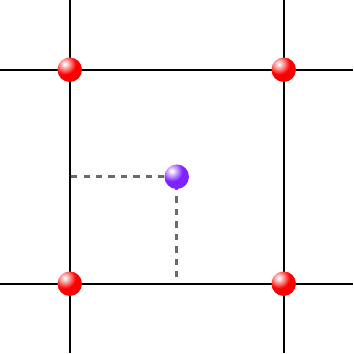}}%
    \put(0.23794324,0.11351729){\color[rgb]{0.49411765,0.13333333,1}\makebox(0,0)[lt]{\lineheight{1.25}\smash{\begin{tabular}[t]{l}$a$\end{tabular}}}}%
    \put(0.52904289,0.4038716){\color[rgb]{0.49411765,0.13333333,1}\makebox(0,0)[lt]{\lineheight{1.25}\smash{\begin{tabular}[t]{l}$b, c$\end{tabular}}}}%
    \put(0.4826066,0.09652905){\color[rgb]{0.49411765,0.13333333,1}\makebox(0,0)[lt]{\lineheight{1.25}\smash{\begin{tabular}[t]{l}$\frac{1}{2}$\end{tabular}}}}%
    \put(0.1229879,0.4805262357){\color[rgb]{0.49411765,0.13333333,1}\makebox(0,0)[lt]{\lineheight{1.25}\smash{\begin{tabular}[t]{l}$\frac{1}{2}$\end{tabular}}}}%
  \end{picture}%
\endgroup%

\caption{Positions of orbitals in the unit cell.}
\label{fig:LatticeLayout}
\end{figure}

The system we consider is a square lattice with an $a$ orbital positioned at each lattice vertex, and the orbitals $b,c$ at a position of $(1/2, 1/2)$ relative to $a$ (see fig. \ref{fig:LatticeLayout}).
The interaction matrix for the Euler CSL model in Eq.~(\ref{eq:Hamiltonian_m}) with tuning parameter $m$ corresponds to the real-space Hamiltonian
\begin{align}
    H_m &= \sum_{\vb r \in \text{u.c.}} \left( \left[ \mathcal{C}_m^3(\vb r) \right]^2 + \left[ \mathcal{C}^2(\vb r) \right]^2 \right),
\end{align}
where
\begin{subequations}
\begin{align}
\begin{split}
    \mathcal{C}_m^3(\vb r) ={}& 2m \, S^a(\vb r) + S^a(\vb r + \vb e_x) + S^a(\vb r - \vb e_x) + S^b(\vb r - \vb e_y) + S^b(\vb r - \vb e_x) + S^c(\vb r) + S^c(\vb r - \vb e_x - \vb e_y),
\end{split}
    \\
    \mathcal{C}^2(\vb r) ={}& \frac{1}{2} \left( -S^b(\vb r) - S^b(\vb r - \vb e_x - \vb e_y) + S^c(\vb r - \vb e_x) + S^c(\vb r - \vb e_y) \right).
\end{align}
\end{subequations}
Similarly, the real space form of the interpolating interaction matrix Eq.~(\ref{eq:BraidingCoupling}) is given by
\begin{align}
    \tilde{H}_{\epsilon} &= \sum_{\vb r \in \text{u.c.}} \left( \left[ (1-\epsilon) \, \mathcal{C}_{m_1}^3(\vb r) + \epsilon \, \mathcal{C}^2(\vb r) \right]^2 + \left[ \epsilon \, \mathcal{C}_{m_2}^3(\vb r) + (1-\epsilon) \, \mathcal{C}^2(\vb r) \right]^2 \right).
\end{align}
The form given here interpolates between any two values $m_1$ and $m_2$ of the parameter in the model in Eq.~(\ref{eq:Hamiltonian_m}); in the main text we set $m_1=-1.2$ and $m_2=-0.8$.

\section{Gauss's laws}\label{sec:Gauss}
As discussed in, for example, Refs. \cite{Yan2023typology, Yan2023detailed}, momentum-space pinch points of the Hamiltonian lead to effective Gauss's laws which govern the conserved quantities at low energies. If a (Fourier-transformed) constrainer has a singularity at a point $\vb{k}_0$ in the Brillouin zone, the form of the corresponding Gauss's law may be found by expanding the constrainer around $\vb{k}_0$, and then performing the mapping $\ii q_x\mapsto \partial^{(\vb{k}_0)}_x$ and $\ii q_y\mapsto \partial^{(\vb{k}_0)}_y$ to recover the long-wavelength effective Hamiltonian. Here 
\begin{equation}
    \partial_{\vb{a}}^{(\vb{k}_0)}S(\vb{r}) = S(\vb{r}) - \ee^{\ii \vb{a}\cdot\vb{k}_0} S(\vb{r}-\vb{a})
\end{equation}
is the so-called `phase-shifted derivative', which reduces to the finite difference $S(\vb{r}) - S(\vb{r}-\vb{a})$ when $\vb{k}_0=\vb{0}$ is located at the $\Gamma$ point.

For the model Eq.~(\ref{eq:Hamiltonian_m}), the constrainers are equal to the (unnormalised) eigenvectors $\vb{v}^3_m(\vb{k})$ and $\vb{v}^2(\vb{k})$, which are orthogonal. For all $m$, pinch points between the lowest two bands occur at the points $\pm X=(\pm\pi, 0)$ and $\pm Y=(0, \pm\pi)$ (note that the Brillouin zone has diameter $4\pi$). Expanding $\vb{v}^2(\vb{k})$ to leading order about these points we find 
\begin{subequations}
\begin{align}
    \vb{v}^2(\pm X+\vb{q}) ={}& \pm\frac{1}{2}\big(0, q_x+q_y, -q_x +q_y\big)^\intercal +\mathcal{O}(\vb{q}^2), \\
    \vb{v}^2(\pm Y+\vb{q}) ={}& \pm\frac{1}{2}\big(0, q_x+q_y, + q_x - q_y\big)^\intercal +\mathcal{O}(\vb{q}^2),
\end{align}
\end{subequations}
corresponding to the real-space Gauss's laws
\begin{subequations}
\begin{align}
    \partial_x^{(X)} S_- +\partial^{(X)}_y S_+ ={}& 0 \\ 
    \partial_x^{(Y)} S_+ +\partial^{(Y)}_y S_- ={}& 0 
\end{align}
\end{subequations}
where $S_\pm(\vb{r}) = S^b(\vb{r})\pm S^c(\vb{r})$, and we have noted that $\partial^{(X)}=\partial^{(-X)}$ and $\partial^{(Y)} = \partial^{(-Y)}$. Note that these conservation laws are independent of the parameter $m$. Since $m$ determines the value of the Euler class $\chi$, this implies that the emergent Gauss's laws are in general independent of the Euler class of the system.

In addition to the pinch points described above, at the band closing points $m=\pm 1$ the top band touches the flat bottom band, creating a pair of second-order pinch points where all three bands coincide. This leads to an additional set of conservation laws deriving from the third band. Again expanding around $\pm X$ and $\pm Y$, we find
\begin{subequations}
\begin{align}
    \vb{v}^3_{m}(\pm X+\vb{q}) ={}& \big(2(m-1) +q_x^2, 0, 0\big)^\intercal \pm \big(0, -q_x+q_y, -q_x -q_y\big)^\intercal + \mathcal{O}(\vb{q}^2),\\
    \vb{v}^3_{m}(\pm Y+\vb{q}) ={}& \big(2(m+1) - q_x^2, 0, 0\big)^\intercal \pm \big(0, +q_x-q_y, -q_x -q_y\big)^\intercal + \mathcal{O}(\vb{q}^2).
\end{align}
\end{subequations}
Hence, at $m=+1$ the band touching at $\pm X$ generates the independent Gauss's laws $\big(\partial_x^{(X)}\big)^2 S^a$ and  $\partial_x^{(X)} S_+ = \partial_y^{(X)}S_-$, while for $m=-1$ we have $\big(\partial_x^{(Y)}\big)^2 S^a$ and $\partial_x^{(Y)} S_- = \partial_y^{(Y)}S_+$. We see here that the additional band-touching point is essentially trivial: The scalar $S^a$'s Gauss's law decouples from that of $S_\pm$.

\end{document}